\documentclass[10pt,preprint]{iopart}
\usepackage{epsfig}

\begin{document}

\title{Localization properties of vibrational modes in $a$-Si$_3$N$_4$} 

\author{Luigi Giacomazzi$^{1,2}$}

\address{ $^{1}$ SISSA, Scuola Internazionale Superiore di Studi Avanzati, 
        via Bonomea 265, I-34136 Trieste, Italy.}
\address{$^{2}$ Materials Research Laboratory, University of Nova Gorica, 
Vipavska 11c 5270-Ajdov\v{s}\v{c}ina, Slovenija.}

\eads{\mailto{giacomaz@sissa.it}}

\date{\today}

\begin{abstract}
We present a first-principles investigation of the localization properties of 
vibrational modes in amorphous silicon nitride ($a$-Si$_3$N$_4$).
Our investigation further confirms that the vibrational modes underlaying the 
peak at $\sim$471 cm$^{-1}$ in the infrared spectrum of silicon nitride 
mainly consist of nitrogen motion in the direction normal to the 
nearest Si neighbors plane. 
In-plane stretching of N--Si bonds becomes largely dominant above $\sim$700  
cm$^{-1}$. In particular vibrational modes underlaying the infrared 
peak at 825 cm$^{-1}$ arise from  Si--N bonds stretching motions. 
If N--N homopolar bonds were present, we show that 
N--N bond stretching  occur above $\sim$1100 cm$^{-1}$. 
Furthermore, we investigate the localization properties of vibrational modes
by calculating their inverse participation ratio (IPR) and phase quotient. 
From this analysis we infer that modes above 600 cm$^{-1}$  shows a 
progressive increase of the localization degree and optic-like behavior, 
especially above 1000 cm$^{-1}$.
At about 650 cm$^{-1}$, given the considerable IPR value, and 
on the basis of projectional anlysis on 
silicon-breathing-like motions of the NSi$_3$ units, 
we suggest that vibrational modes 
may involve correlated motions of neighboring SiN$_4$ tetrahedra. 

\end{abstract}



\section{Introduction}
%
%

Nowadays amorphous silicon nitride ($a$-Si$_3$N$_4$) is widely used  
in  microelectronics e.g. to fabricate insulating layers in triple oxide-nitride-oxide 
structures \cite{ONOs}.
In particular, because of its high concentration of charge traps,
$a$-Si$_3$N$_4$ is employed as charge storage layer in nonvolatile 
memory devices \cite{MNOS,Vianello2011}. 
Thin films of $a$-Si$_3$N$_4$ are grown either through chemical vapor deposition \cite{HandbookCVD} (CVD)
of silane and ammonia gases or through physical vapor deposition (PVD) \cite{HandbookPVD}.
Infrared spectroscopy can then be adopted for characterizing films composition and 
stoichiometry \cite{Debieu2013}. 
For instance, the intensity of infrared absorption peaks around 2220 and
3300 cm$^{-1}$, arising from Si--H and N--H stretching modes respectively, 
can be used to estimate the hydrogen content \cite{Luc83,Ver06,LinLee}.

%
%
The experimental infrared dielectric function of $a$-Si$_3$N$_4$ features 
two main broad peaks at 471 cm$^{-1}$ and 825 cm$^{-1}$ 
\cite{Tice2010,Gunde01}.
These peaks are then slightly shifted in the absorption spectrum where they 
appear respectively at about 490 cm$^{-1}$ and 900 cm$^{-1}$ \cite{Tice2010}.
The former peak was attributed to silicon-breathing motion \cite{Luc83,NotaTri},
meanwhile the latter high frequency peak at 900 cm$^{-1}$ is commonly 
assigned to Si--N bond stretching \cite{Luc83,Pandey04}. 
The assignement of the  490 cm$^{-1}$ feature \cite{Luc83} invoked a breaking 
of symmetry caused by the presence of hydrogen atoms. However as shown in 
Ref.\ \cite{GU09} 
the presence of hydrogen seems not very relevant for explaining this
feature, that instead appear to be originated by out-of-plane nitrogen motion.
In Ref.\ \cite{Yin90} a feature at about 640  cm$^{-1}$ 
was observed in the absorption spectrum of silicon nitride films. 
In particular in  Ref.\ \cite{Yin90} it is shown that this
feature is not affected by hydrogen content and thus can not 
be related to Si--H stretching modes. 
By invoking an analogy to the absorption bands in $a$-SiO$_2$, 
this feature at 640 cm$^{-1}$  was tentatively
assigned to some unspecified vibration of the NSi$_3$ unit \cite{Yin90}.
Moreover it has been proposed that a feature
appearing at  644 cm$^{-1}$ could be originated by silicon-breathing-like
motions of the NSi$_3$ unit \cite{Tice2010}. Incidentally we note that such 
an hypothesis suggests implicitly a larger localization degree of the vibrational 
modes with respect to the typical localization of modes in glasses up to 
a few hundred cm$^{-1}$. 
On the other hand, in recent years the vibrations in the medium frequency range (520 to 770 
cm$^{-1}$) were further ascribed, for the $\alpha$-Si$_3$N$_4$ crystalline phase, 
to vibrations of Si atoms in SiN$_4$ tetrahedra \cite{Leg2014,Kuzuba1978}. 


%
%
On the theoretical side, density functional approaches have 
successfully complemented and helped to interpret experimental 
spectroscopic data, thereby considerably enriching the knowledge 
of the network organization and ions dynamics in several
materials e.g.  v-SiO$_2$ \cite{LG09,PC97,PC98}, v-GeSe$_2$ \cite{LG2011,Kali2013}, 
v-B$_2$O$_3$  \cite{Puma05}.
Despite in the recent years several theoretical works have investigated 
the structural and electronic structure properties of systems based on 
silicon nitride \cite{Pham2011,Pham2013,Hin2013,NLA2015,Das2018}, only 
a few of them have discussed the vibrational properties \cite{Leg2014,Xu2011}.

%
%
In this paper, we discuss the origin of vibrational bands in
silicon nitride by comparing the vibrational spectra of two first-principles 
model structures of amorphous silicon nitride. 
The first of these two models was generated in Ref. \cite{GU09} and contains 
a small amount of hydrogen atoms ($\sim$0.1\% wt).  
%
%
A second smaller model, hydrogen free, is generated in this work with 
the aim of strengthening our analysis 
by exploiting comparisons among structurally different models, 
by ruling out any hypothetical relevant effect due to the hydrogen 
presence on the v-DOS in the medium frequency range, 
and finally by allowing to discuss specific vibrations of 
N-N homopolar bonds
which are absent in our first model \cite{GU09}.    
The vibrational density of states are first analysed by means of 
decomposition into N and Si contributions.
Then, silicon nitride vibrational modes are analysed in terms of vibrations 
of the N--N and NSi$_3$ units. 
Finally we investigate the nature of vibrational modes by 
examining their localization properties through the inverse participation
ratio (IPR) and phase quotient \cite{Taraskin}. 
%
The present analysis aims at giving a general description of the
vibrational modes of $a$-Si$_3$N$_4$ with a special attention to those 
around $\sim$650 cm$^{-1}$, at the crossover between the two main absorption bands,
Moreover, this investigation further discusses the assignment of the 
peak at 490 cm$^{-1}$ in the infrared absorption spectrum to 
vibrational modes featuring  nitrogen motion 
in direction normal to the silicon neighbors plane \cite{Kuwa2008,GU09},
by considering analysis on multiple models.

%
%
The paper is organised as follows.
In Sec. \ref{Sec_Methods}, we concisely describe the methods adopted in the present work. We
also outline how we generated our
model structures as well as their main structural features. 
In Sec. \ref{Sec_Vib}, we analyse the vibrational density of states 
and the localization aspects of the vibrational modes of our models. 
Furthermore, we discuss the infrared  spectra of  our models in 
comparison with experimental results.
We draw conclusions in Sec. \ref{Sec_Concl}.

%
%
\section{Methods and models}\label{Sec_Methods}
\subsection{Methods}
In the present work, we have performed first-principles electronic-structure calculations
based on density functional theory. For generating the model structures of
$a$-Si$_3$N$_4$ we carried out molecular dynamics simulations
using the Car-Parrinello method \cite{CarPar,Pasquarello92}.
In particular, we used the computational codes and pseudopotentials
from the   {\sc Quantum-Espresso} package \cite{qe}.
The exchange and correlation functional was approximated through
the local density  approximation (LDA) \cite{pz81}. 
Core-valence interactions were described through ultrasoft pseudopotentials \cite{Vanderbilt} 
for N  and through a normconserving pseudopotential for Si atoms. 
The electronic wavefunctions and the charge density were expanded 
using plane waves basis sets defined by energy cutoffs of
25 Ry and 200 Ry, respectively.
The Brillouin zone was sampled at the $\Gamma$ point.
We derived the vibrational frequencies and eigenmodes by diagonalizing
the dynamical matrix, \cite{LG09} which we calculated numerically by taking finite
differences of the atomic forces.
To this purpose we used atomic displacements of 0.05 Bohr.
%
%
In Ref. \cite{GU09} we obtained the relevant coupling tensors 
needed to calculate infrared spectra by taking advantage of a
scheme for  applying a finite electric field in periodic 
cell calculations \cite{PumaFields,LG09}. 
An alternative way for calculating the coupling tensors i.e
the dynamical charges is to perform linear-response calculations 
\cite{Baroni} as implemented in the {\sc Quantum-Espresso} package \cite{qe}.
In this work we obtained the dynamical charges of the model
labelled as Model II (see \ref{ModelGen})
by following the latter approach.

\subsection{Model generation procedure}\label{ModelGen}
%
%
In this work we consider two models of silicon nitride. The first one 
was generated by first-principles molecular dynamics \cite{GU09} through a
quench-from-the-melt approach, as described here above. 
This model has a size of 152 atoms in a simulation cell of 22.1 bohr. 
In the following we will refer to this model as Model I. 
A smaller model with a size of 70 atoms  (Model II) was here generated 
starting from a supercell of $\beta$-Si$_3$N$_4$
that was rescaled to match  the experimental value of the density
3.1 g/cm$^{3}$ \cite{Sze,Supercell}.
Molecular dynamics runs were then performed for obtaining the model of
$a$-Si$_3$N$_4$. First the system was thermalized at the temperature of 3700 K for 12 ps
using a Nos\'e-Hoover thermostat \cite{nose}. Successively, the sample
was quenched for 7 ps down to 1600 K below the theoretical
melting point. Finally, the structural geometry was further
optimized by a damped molecular-dynamics run.
%
%
\begin{figure}
 \begin{center}
  \includegraphics[width=7.5 cm]{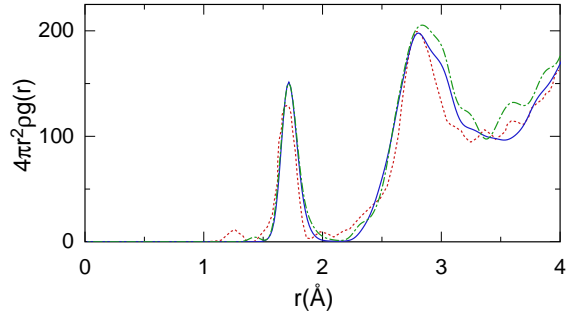}
 \end{center}
\caption{\label{fig_rdf}
Radial distribution function of a-Si$_3$N$_4$ calculated at room
temperature for Model I (solid/blue), Model II (dot-dashed/green)  and experimental
data from Ref.\ \protect\cite{Misawa79} (dotted/red).
}
\end{figure}

%
%
%
\subsection{Structural properties}
In Tab.\ \ref{tabstr} we summarize the main structural parameters 
of our models. 
Both Model I and II show an average Si--N bond length of about 1.73 \AA\, 
with a similar standard deviation of $\sim$0.06 \AA. 
The average Si-N-Si and N-Si-N  bond angles are 117$^{\circ}$ and 109$^{\circ}$ as typical for 
silicon nitride models \cite{Jarolimek,Vash95}. 
The calculated radial distribution function of Model II [Fig.\ \ref{fig_rdf}] 
is very similar to that found for Model I  \cite{GU09} and in fair agreement 
with the experimental results \cite{Misawa79}, thus assuring that also
Model II provides a fair structural description of silicon nitride. 
In Model II we register the presence of ten two-fold rings, while
in Model I we found sixteen \cite{GU09}.
47\% of silicon atoms in Model II belong to such rings, slightly more than
in Model I where we register  39\%  \cite{GU09}.
A few under- and over-coordinated nitrogen and silicon atoms are also 
found. Moreover the structure contains a N--N homopolar bond  with 
a bond length of 1.4 \AA. This is rather close to the value ($\sim$1.2 \AA\,) reported 
e.g. for the N--N bond length in a first-principle investigation of the 
liquid phase \cite{Mauri2011} and is 0.2 \AA\, shorter than the value found in \cite{Hin2012}
for a nonstoichiometric system Si$_3$N$_{4.5}$. 
We note however that the concentration of such type of bonds in $a$-Si$_3$N$_4$
is rather controversial \cite{Hin2012,Ipp2011,Kroll01} and investigating 
this issue goes beyond the scope of the present work.
Despite the presence of point defects as mentioned above,
in both Model I and II  most ($\sim$ 90\%) of nitrogen and silicon  atoms 
are three- and four-fold coordinated \cite{GU09}.  

%
%
\begin{table}
\caption{\label{tabstr}
Structural parameters of our models of $a$-Si$_3$N$_4$. 
$N$ is the number of atoms in the simulation cell.  
$d_{\rm SiN}$ indicates the bond length and is given in \AA.
Si-N-Si  corresponds to the average intertetrahedral angle.
Standard deviations are given in parenthesis. 
Experimental $d_{\rm SiN}$ is taken froms Ref.\ \protect\cite{Misawa79} and 
values referring to Model I are taken from Ref.\ \protect\cite{GU09}.
}

\begin{center}
 \begin{tabular}{lllll}
     & $N$ & $d_{\rm SiN}$   & N-Si-N & Si-N-Si   \\
    
 \hline
Model  I & 152          & 1.73 (0.06)   & 109.1$^{\circ}$ (13$^{\circ}$)  & 117.2$^{\circ}$
(15.1$^{\circ}$) \\
Model II & 70           & 1.74 (0.07)   & 108.9$^{\circ}$ (13.4$^{\circ}$) & 
117.0$^{\circ}$ (15.9$^{\circ}$) \\
Expt.   &   & 1.729         &              &          \\
\hline
\end{tabular}
 \end{center}
\end{table}

%
%
\section{Vibrational density of states}\label{Sec_Vib}

\subsection{Nitrogen out-of-plane and stretching  motions} 
\begin{figure}
 \begin{center}
  \includegraphics[width=7.5 cm]{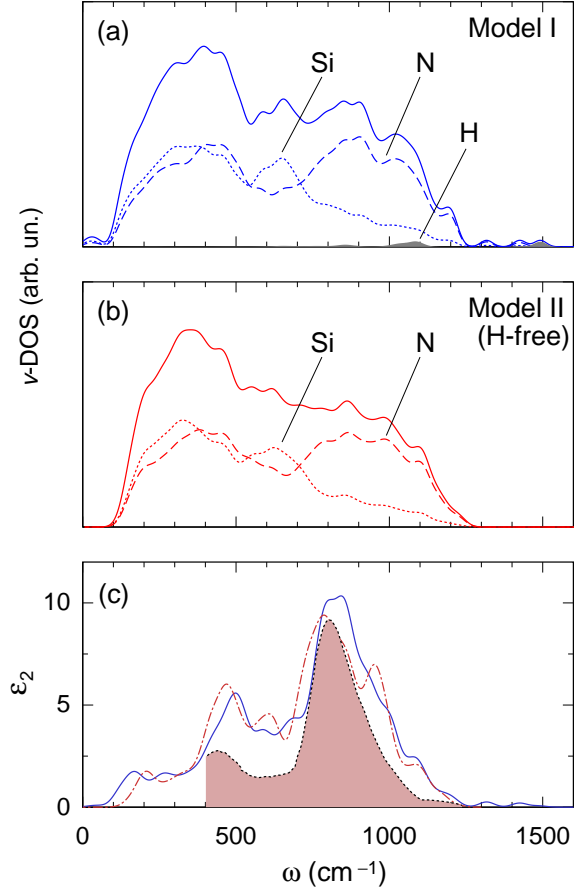}
 \end{center}
\caption{\label{vdos-m70}
 v-DOS (solid) and silicon (dotted) and nitrogen (dashed) contribution to the
 density of states for (a) Model I [Ref.\ \protect\cite{GU09}]
 and (b) Model II. 
(c)
Calculated imaginary part of the dielectric function ($\epsilon_2$) 
for Model I (solid/blue) and for Model II (dot-dashed/red) 
and experimental results (dotted/shaded pink) of Ref.\ \protect\cite{Gunde01}.
A Gaussian broadening of 25 cm$^{-1}$ was applied. 
}
\end{figure}

In Fig.\ \ref{vdos-m70} we compare the v-DOS of the two models of $a$-Si$_3$N$_4$. 
The v-DOS of the two models are quite similar with a pronounced peak at about 
400 cm$^{-1}$ and broad bands extending up to about 1250 cm$^{-1}$.
%
%
Next, in  Fig.\ \ref{vdos-m70}(c), we show the imaginary part of the 
dielectric function of Model I and II. 
Both models  show infrared spectra of similar fair quality when compared 
to the  infrared spectrum  obtained from experiments \cite{Gunde01}. 
Minor differences between the two spectra of Model I and II are noticeable
especially in the central frequency region from about $\sim$500 cm$^{-1}$ up to
900 cm$^{-1}$. These minor differences should be attributed to small structural 
differences (Fig.\ \ref{fig_rdf}) between the two models as well as to size effects
and are quite typical in modelling glassy materials \cite{LG06,LG2011,LG09}. 

The decomposition of the v-DOS shown  in Fig.\ \ref{vdos-m70}(a) 
for Model I,  in terms of silicon and nitrogen and hydrogen weights 
is essentialy the same  as that shown in Fig.\ \ref{vdos-m70}(b) 
for the v-DOS of Model II, thus further supporting the reasonable 
assumption that a low concentration of hydrogen will provide 
only a negligible contribution to the v-DOS below $\sim$1000 cm$^{-1}$
and thus that Model I is perfectly suitable to discuss the vibrational 
properties also of pure Si$_3$N$_4$ glass. 
In both models the nitrogen contribution to the v-DOS features a 
two humps camel shape with broad maxima at about 400 and 900 cm$^{-1}$, 
while the silicon contribution has broad maxima at about 350 and 650 cm$^{-1}$.
Despite in  Fig.\ \ref{vdos-m70}(a) the peak at 650 cm$^{-1}$ of 
the silicon contribution is more pronounced than the corresponding one in 
Fig.\ \ref{vdos-m70}(b), we note that for both models we can easily identify 
such a feature.
We thus conclude that this peak at about 650 cm$^{-1}$ in the v-DOS 
does constitute a distinctive feature of the v-DOS of glassy silicon nitride 
as well as of the crystalline phase $\alpha$-Si$_3$N$_4$ where the crystal 
symmetries appear to enhance it \cite{Leg2014}.
By contrast, we note that such a feature is absent in the v-DOS obtained 
by using force fields for models generated by classical molecular 
dynamics \cite{Vash95,Omel96,Loong95}.  
Classical modeling approaches are generally not sufficiently
accurate \cite{Sarnt97,BK02} for describing vibrational properties of vitreous
silica, thus we suggest that classical force fields 
also fail in describing silicon nitride vibrational modes around 650 cm$^{-1}$. 

%
%
%
In Fig.\ \ref{vdos-sib}(a) we decompose the nitrogen contribution to the
v-DOS of Model II according to three orthogonal directions defining the
local environment of the N atoms  as we have previously done 
for Model I \cite{GU09}. 
The first direction,
which we refer to as out-of-plane or rocking, is taken orthogonal 
to the plane of the three nearest Si neighbors. Then, the second one is taken along
the bisector of one of the three Si-Si-Si angles and the third
one is given by the cross product of the first two. The latter
two directions define the stretching motion in the plane of
the three nearest Si neighbors. 
This decomposition in terms of rocking and stretching motions 
allow to establish the existence of two separate
nitrogen bands below and above $\sim$600 cm$^{-1}$.
Fig.\ \ref{vdos-sib}(a) further supports the result of Ref.\ \cite{GU09} 
i.e. the contribution to the
nitrogen partial v-DOS due to out-of-plane motions is maximised at about
350 to 450 cm$^{-1}$, while the contribution due to stretching motions
accounts for almost all the nitrogen  v-DOS above $\sim$700 cm$^{-1}$. 

\subsection{Projection on NSi$_3$ silicon-breathing and N--N stretching modes}
\begin{figure}
 \begin{center}
  \includegraphics[width=7.5 cm]{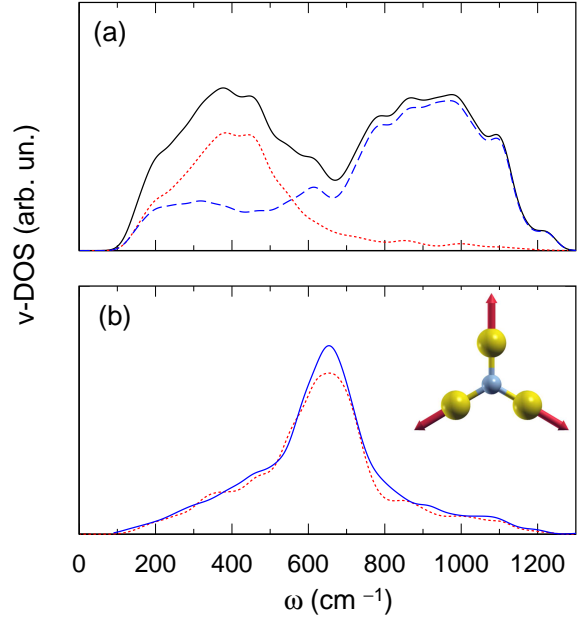}
 \end{center}
\caption{\label{vdos-sib}
(a) Decomposition of the nitrogen contribution (solid) to the
v-DOS of Model II into motions along the normal 
to the plane of silicon neighbors (dotted/red) and into stretching 
motions in the plane of silicon neighbors (dashed/blue). 
Only threefold coordinated nitrogen atoms are taken into account.
(b) Average projections of the vibrational eigenmodes onto silicon-breathing 
motions for Model I (solid/blue) and Model II (dotted/red).
A Gaussian broadening of 25 cm$^{-1}$ was applied. 
}
\end{figure}

%
%
Infrared spectra of amorphous silicon nitride show a small, yet non-negligible
intensity between the two main peaks at 471 and 825 cm$^{-1}$. 
In Ref.\ \cite{Tice2010,Yin90} it was proposed that modes belonging to this
central region, and in particular a fitted absorption band at 
$\sim$640 cm$^{-1}$, could be interpreted in terms of some vibration of the
NSi$_3$ unit, eventually in terms of the silicon-breathing mode [Fig.\
\ref{vdos-sib}(b)].
%
%
For symmetry reasons this mode of the NSi$_3$ unit is Raman active 
but not infrared active \cite{Luc83}. 
However the ideal symmetry shown in the mode depicted in Fig.\ \ref{vdos-sib}(b) 
is not necessarily respected in the silicon nitride network where 
 NSi$_3$ units may have a distorted geometry. 
As soon as the symmetry is broken the silicon-breathing mode can become 
infrared active \cite{Luc83}. Moreover from Figs.\ \ref{vdos-m70} and \ref{vdos-sib} 
we remark that at about 650 cm$^{-1}$ nitrogen displacements do 
stretch N--Si bonds, so that in general we do not have just silicon 
displacements as in the very ideal case of Fig.\ \ref{vdos-sib}(b), but 
more in general vibrational modes will involve several SiN$_4$ tetrahedra,
and hence a considerable nitrogen motion. 
Still, it is interesting to investigate what type of 
silicon motion may take place in the frequency range between 
the two main peaks of the infrared dielectric function and 
if the projection on some NSi$_3$ unit mode can be of help to 
understand some property of the vibrational modes in the 
region 500 to 700 cm$^{-1}$. 
%
%
By considering the vibrational modes of  our model structures
we investigate the frequency localization of silicon-breathing motions 
in amorphous silicon nitride.
First, we project the vibrational eigenmodes onto the ideal silicon-breathing 
mode [Fig.\ \ref{vdos-sib}(b)] for each NSi$_3$ unit. 
Next by calculating the average of these projections 
we obtain the average weight of silicon-breathing motions 
as function of the frequency \cite{LG06}.
The result shown in Fig.\ \ref{vdos-sib}(b) indicates the presence of just one
broad peak at about 
650 cm$^{-1}$ in both in Model I and Model II
and with a width at half maximum of about two hundreds of cm$^{-1}$.
Furthermore, we explicitly count all the NSi$_3$ units 
showing silicon motion compatible with the ideal silicon-breathing 
mode. In the frequency range between 620 and 680 cm$^{-1}$
silicon-breathing-like motions involve up to 20\% of the NSi$_3$ units. 
The result of Fig.\ \ref{vdos-sib}(b) indicates the presence of correlated 
motions involving a few SiN$_4$ neighboring tetrahedra which at about  
650 cm$^{-1}$ has an important component on the NSi$_3$ plane. 
By comparing with Fig.\ \ref{vdos-m70}(b) we infer that such type of
correlated motion is almost absent below $\sim$400 cm$^{-1}$. 
On the other hand above $\sim$750 cm$^{-1}$  nitrogen 
stretching, involving a higher localization degree as discussed 
hereafter \cite{Nota_Sib_N}, appear to dominate the v-DOS, 
while Si contribution to the v-DOS 
fades away [Fig.\ \ref{vdos-m70}].


\begin{figure}
 \begin{center}
  \includegraphics[width=7.5 cm]{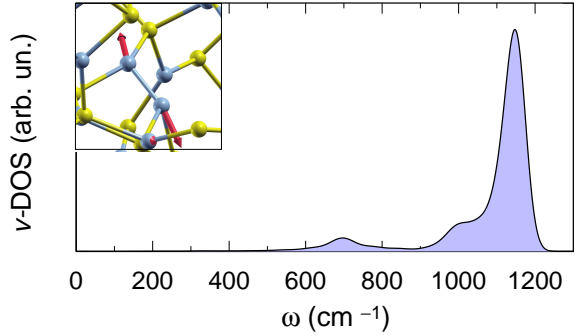}
 \end{center}
\caption{\label{vdos-NN}
Projection of the vibrational eigenmodes of Model II onto 
nitrogen-nitrogen stretching motion. In the inset a snapshot
of a vibrational mode at 1160 cm$^{-1}$ is shown.  
A Gaussian broadening of 25 cm$^{-1}$ was applied. 
}
\end{figure}

In Fig.\ \ref{vdos-NN} we show the vibrational density of states 
obtained by projecting the vibrational eigenmodes on the 
nitrogen-nitrogen stretching motion of the N--N homopolar 
bond present only in Model II. The projection shows a sharp peak at about
$\sim$1150 cm$^{-1}$. We remark that shorter N--N bonds \cite{Ipp2011,Mauri2011}
should give rise to higher frequencies peaks as it is the case for 
instance of two-fold coordinated nitrogen atoms \cite{GU09},
and viceversa longer N--N bonds will show lower stretching frequencies.  
We thus infer that stretching of N--N homopolar bonds should  only affect 
the high frequency tail (above $\sim$1100 cm$^{-1}$)  of the infrared dielectric
function shown in Fig.\ \ref{vdos-m70}(c).
 
%
%

%
%
\subsection{Localization properties of vibrational modes}
\begin{figure}
 \begin{center}
  \includegraphics[width=7.5 cm]{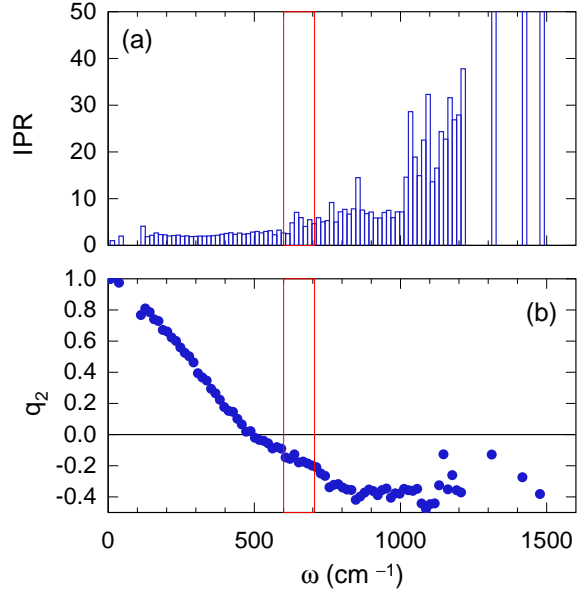}
 \end{center}
\caption{\label{Fig_pq2}
(a) Inverse participation ratio of the vibrational modes of Model I. 
Values are averaged over an interval of 15 cm$^{-1}$. 
(b)
Phase quotient of the vibrational modes of the silicon nitride Model I.
Values are averaged over an interval of 15 cm$^{-1}$. 
Interval 600-700 cm$^{-1}$ is emphasized with a pink box. 
}
\end{figure}

We investigate the spatial localization of the eigenmodes by calculating
the inverse participation ratio (IPR) $\pi_j$ \cite{Taraskin} for the 
eigenmode $j$: 
\begin{equation}
\pi_j=N \frac{\sum_{i=1}^{N} |{\rm \bf u}^j_i|^4}{(\sum_{i=1}^{N}|{\rm \bf u}^j_i|^2)^2}
\end{equation}
where $N$ is the number of atoms, and ${\rm \bf u}^j_{i}$ indicates the diplacement 
of the $i$ atom \cite{LG09}.
For an ideally delocalised mode, the IPR is exactly one, and the larger 
the IPR the more localised the mode is.
In Fig.\ \ref{Fig_pq2}(a) we show the IPR of the vibrational modes 
of our largest model [Model I, Ref.\ \cite{GU09}]. 
We note that modes up to 600--620 cm$^{-1}$ have a quite low IPR 
(up to $\sim$2)
indicating a delocalised nature of these modes. 
However for vibrational modes with frequencies  about $\sim$620 cm$^{-1}$  
we observe an  IPR  of $\sim$6, a few times larger than 
the one of delocalised modes.
Then up to 1050 cm$^{-1}$ IPR shows a slowly increasing trend.
Above 1050 cm$^{-1}$ the IPR further increases to values that indicate
an even stronger localization degree (IPR $\sim$ 30). 
Incidentally we note that the upper limit of the delocalised modes ($\sim$600 cm$^{-1}$) actually 
corresponds to the fading away of the "rocking" modes 
of the nitrogen \cite{GU09}.
%
%
In other words as soon as nitrogen start moving mainly on 
the Si--N bonds plane, then the vibrational modes become slightly 
more localised, and in particular the IPR registers a 
localization peak at about $\sim$650 cm$^{-1}$ 
in correspondance of the projection peak shown in Fig.\ \ref{vdos-sib}. 
The localization then progressively increases and above $\sim$1000 cm$^{-1}$
it shows a jump to $\sim$30 indicating localization of a few 
nitrogen atoms [Fig.\ \ref{vdos-m70}(a)]. 
We note that the slow passage from delocalised to localised states that 
takes place in the frequency window 600-700 cm$^{-1}$ includes the region 
of the peak of the silicon
partial density of states (Fig.\ \ref{vdos-m70}) and also the peak of the projection on the 
silicon breathing mode [Fig. \ref{vdos-sib}(b)].  

Besides the IPR another quantity that is useful for analysing 
the local behavior of the eigenmodes is the phase quotient \cite{Taraskin},
\begin{equation}
q^{(2)}_{j}=
\frac{1}{\sum_{i,i'}|{{\rm \bf e}^j}_i|^2} \sum_i\sum_{i'}
\frac{{\bf u}^j_i\cdot{\bf u}^j_{i'}}{|{\bf u}^j_i|\cdot |{\bf u}^j_{i'} |}
|{{\rm \bf e}^j}_i|^2
\end{equation}  
where ${\bf u}^j_{i}$ indicates the diplacement of the $i$ silicon and 
$i'$ runs over its nearest nitrogen neighbors. $j$ labels the normalised vibrational 
mode ${\rm \bf e}^j$ \cite{LG09}.  
The phase quotient corresponds, for a given silicon atom, to the average 
cosinus of the angle between the atomic displacements of its nearest nitrogen  neighbors.  
In Fig.\ \ref{Fig_pq2}(b) we show the phase quotient ($q_2$) 
calculated for Model I.
Modes at low frequencies   
show an acoustic-like behavior with  neighboring atoms 
moving in phase ($q_2 \sim 1$). 
The phase quotient then decreases almost linearly and at about
500 cm$^{-1}$ becomes zero, indicating that nitrogen and silicon 
atoms have on average orthogonal displacements. Moreover in this region 
[Fig.\ \ref{vdos-m70}(a) and (b)] nitrogen rocking motions are dominant. 
Above $\sim$600 cm$^{-1}$ modes start to show an optic-like behavior
with neighbors moving in anti-phase ($-1<q_2 <0$). The phase quotient further
decreases in the stretching region above 800 cm$^{-1}$. 
The behavior of the phase quotient in silicon nitride appears qualitatively
similar to that registered in v-SiO$_2$ with an almost linear trend
up to 800 cm$^{-1}$ where we note a saturation $q_2 \sim -0.4$ corresponding
to the Si--N bond stretching modes \cite{Taraskin}. 
 
%
%

%
%
\section{Conclusion}\label{Sec_Concl}
The present first-principles investigation 
has focussed on localization aspects of vibrational modes in $a$-Si$_3$N$_4$ systems. 
Moreover, the present work provides further support for the 
interpretation of the 471 cm$^{-1}$ and 825 cm$^{-1}$ peaks of the
infrared dielectric function of $a$-Si$_3$N$_4$ as due to 
nitrogen motion in the direction normal to the nearest Si neighbors plane 
and to Si--N bond stretching motion, respectively \cite{GU09}. 
%
From the inverse participation ratio and phase quotient analysis
we infer that modes above 600 cm$^{-1}$ show a progressive increase
of localization and optic-like behavior with modes that become highly localized
 above 1000 cm$^{-1}$. 
In particular at about 650 cm$^{-1}$ the vibrational modes are considerably 
more localized than below 600 cm$^{-1}$ and,
moreover, our analysis based on projection onto silicon-breathing-like modes 
of the NSi$_3$ units suggests that correlated motions of a few neighboring SiN$_4$ 
tetrahedra can take place at about 650 cm$^{-1}$, 
and may sometimes appear as ring modes, 
at the crossover frequency between the two main 
bands of the infrared spectrum. 

%
%

%

\section*{References}


\begin{thebibliography}{10}
\bibitem{NotaTri}
In molecular systems such as trisilylamine, a mode at 490 cm$^{-1}$ 
is assigned to a silicon-breathing vibration \cite{Trisyl}. 
The 471 cm$^{-1}$ frequency peak of the infrared spectrum of $a$-Si$_3$N$_4$
is clearly quite close to the frequency of the  silicon-breathing vibration
in  trisilylamine.
However in passing from a molecular system to an extended one some caution
should be used and in general there is no strict correspondence between 
the vibrational modes of an extended system and 
those of a molecule formed by atoms of the same species.
\bibitem{Supercell}
We used a 2x2x2 $\beta$-Si$_3$N$_4$ supercell that was reshaped from hexagonal to 
tetragonal without changing the density. 
Next, we further reshaped the supercell into a cubic one matching the 
experimental density.
\bibitem{Nota_Sib_N}
A large nitrogen diplacement is clearly not much compatible 
with the local NSi$_3$ mode of Fig.\ \ref{vdos-sib}, and this 
is reflected by the decrease of the projection above 700 cm$^{-1}$.
\end{thebibliography}
\end{document}